\documentclass[letterpaper,twocolumn,english,groupedaddress,superscriptaddress,showpacs]{revtex4}
\usepackage{ae}
\usepackage{aecompl}
\usepackage[T1]{fontenc}
\usepackage[latin1]{inputenc}
\usepackage{amsmath}
\usepackage{graphicx}
\usepackage{amssymb}
\usepackage{amsthm}
\usepackage{babel}

\begin{document}

\title{The Double Exchange Model at Low Densities}

\date{\today}



\author{Vitor~M.~Pereira}

\affiliation{Department of Physics, Boston University, 590 Commonwealth Avenue,
Boston, Massachusetts 02215, USA}

\affiliation{CFP and Departamento de F{\'\i}sica, Faculdade de Ci{\^e}ncias Universidade
do Porto, 4169-007 Porto, Portugal.}

\author{J.~M.~B.~Lopes~dos~Santos}

\affiliation{CFP and Departamento de F{\'\i}sica, Faculdade de Ci{\^e}ncias Universidade
do Porto, 4169-007 Porto, Portugal.}

\author{A.~H.~Castro~Neto}

\affiliation{Department of Physics, Boston University, 590 Commonwealth Avenue,
Boston, Massachusetts 02215, USA}


\begin{abstract}
We obtain the phase diagram of the double-exchange model at low
electronic densities in the presence of electron-electron interactions.
The single particle problem and its extension to low electronic
densities, when a Wigner crystal of magnetic polarons is generated
due to unscreened Coulomb interactions, is studied. It is argued that the Wigner crystal
is the natural alternative to phase separation when the Coulomb interaction
is taken into account. We discuss the thermal and quantum 
stability of the crystalline phase towards a polaronic Fermi liquid
and a homogeneous, metallic, ferromagnetic phase. The relevance and application of our results to EuB$_6$ is also considered.
\end{abstract}

\maketitle



{\it Introduction}. 
Magnetic polarons are ubiquitous in the physics of double-exchange
systems such as mixed valence manganites \cite{Gennes:1960}, antiferromagnetic \cite{Nagaev:2002} and ferromagnetic semiconductors (such as
EuO \cite{Oliver:1972}), diluted magnetic semiconductors (DMS)
(such as Ga$_{1-x}$Mn$_{x}$As) \cite{Dietl:1982,Kaminski:2002}, and 
in EuB$_6$ where they have been observed in the optical
response \cite{Snow:2001}.
Colossal magnetoresistance (CMR) has been attributed in part to the
presence of magnetic polarons close to the paramagnetic-ferromagnetic
transition \cite{Teresa:2000}. In fact, CMR in the Mn pyrochlores
has been proposed on the basis of magnetic polarons \cite{Majumdar:1998}.
Given that these different classes of magnetic materials have attracted
considerable attention in recent years because of their potential
for the development of new magnetoelectronic devices, a number
of theoretical approaches to the problem have been developed through
the years. In particular, extensive work has been done with emphasis
in the physics of magnetic
semiconductors \cite{Kasuya:1970,Dietl:1982,Heiman:1983,Nagaev:2002} 
and CMR manganites \cite{Varma:1996Batista:2000Garcia:2002Meskine:2004},
where electron-electron interactions are assumed to be weak.  

\begin{figure}
\begin{center}
\includegraphics[width=0.9\columnwidth]{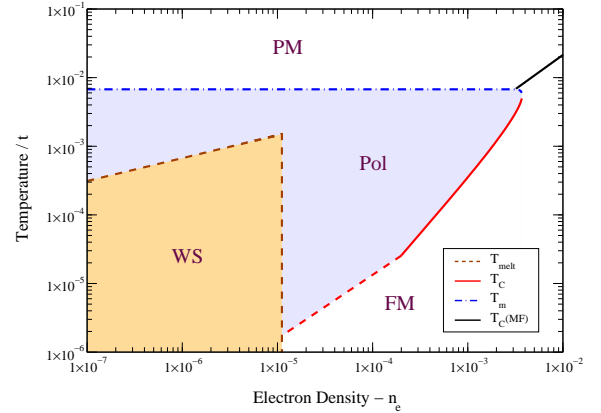}
\end{center}
\caption{Phase diagram of the diluted DEM model as a function
of temperature and density ($t=0.5\,\textrm{eV},a=4\,\textrm{\AA}$):
paramagnetic (PM), polaronic (Pol),  
polaronic Wigner crystal (WS), and ferromagnetic (FM).}
\label{fig:PhDiag_WS_Polarons} 
\end{figure}

In this work we will focus our attention on the phase diagram of the 
double-exchange model (DEM) at low densities,
taking into account the effects of Coulomb interactions between the
electrons. Our results are summarized in Fig.~\ref{fig:PhDiag_WS_Polarons}. 
Studies devoted to the polaronic
stability in this particular model and its variations have been performed
by several authors, both analytically and numerically \cite{Daghofer:2004aDaghofer:2004bKoller:2003Neuber:2005Pathak:2001Wang:1997Yi:2000},
but no unified theory has been presented so far. Here we
study the extreme case of zero density (that is, the single magnetic
polaron), the Wigner crystallization of magnetic polarons due to 
Coulomb interactions, and its thermal and quantum melting
into a polaronic Fermi liquid and a ferromagnetic metal. 
We argue that the phase separation instability of the non-interacting 
DEM at low densities is replaced by Wigner crystallization and a polaronic
Fermi liquid, stabilized by electrostatic interactions.
We also show that the mean-field approach breaks down 
in the low density regime and that electron-electron interactions have to be
taken into account properly. 

{\it The Model}. 
In the presence of the Coulomb term, the DEM reads (we use units such that $\hbar=1=k_B$):
\begin{equation}
\mathcal{H}_{DE}=-\sum_{\left\langle i,j\right\rangle}
t_{ij}a_{i}^{\dag}a_{j}+\textrm{h. c.}+ e^2 \sum_{i>j}
\frac{(n_{i}-n_{e})(n_{j}-n_{e})}{|{\bf r}_i-{\bf r}_j|},
\label{eq:H_DE_Coulomb}
\end{equation}
where $a_i$ ($a_i^{\dag}$) is the annihilation (creation) operator for
an electron on site ${\bf r}_i$ with its spin aligned in the direction
of a classical localized spin 
${\bf S}_{i}=S(\sin\theta_{i}\cos\phi_{i},\sin\theta_{i}\sin\phi_{i},\cos\theta_{i})$, $\theta_i$ and $\phi_i$ being the local Euler angles. In Eq.~(\ref{eq:H_DE_Coulomb}), $t_{ij}=t[\cos\left(\theta_{i}/2\right)\cos\left(\theta_{j}/2\right)+\sin\left(\theta_{i}/2\right)\sin\left(\theta_{j}/2\right)e^{-i(\phi_{i}-\phi_{j})}]$ is the hopping energy, $e$ the electric charge, $n_i = a^{\dag}_i a_i$ the number operator, and $n_e$ the average number of electrons per atom. 
This model can be obtained from an interacting Kondo lattice Hamiltonian
in the limit where the exchange interaction between local and
electron spin, $J_H$, satisfies $J_H \gg t$ \cite{Nagaev:2002}. 

The presence of the full Coulomb interaction in eq.~(\ref{eq:H_DE_Coulomb})
is to be understood in connection with the case of extremely reduced
electron density. In a conventional Fermi liquid
the high density of electrons makes the screening
process very effective, and the effect of the electron-electron interactions
can be absorbed into the renormalization of physical quantities such as
the effective mass. In a very diluted electron gas the Coulomb
interaction cannot be addressed meaningfully in this way. In the language
of the one component plasma, this can be understood with reference
to the dimensionless parameter $r_{s} = r_o/a_0$, where
$r_o = (3/4\pi\rho)^{1/3}$ ($\rho$ being
the volumetric density) is the average distance between electrons, and 
$a_0$ is the Bohr radius. While 
the kinetic energy scales as $1/r_{s}^{2}$, the potential
energy varies as $1/r_{s}$, and, therefore dominates in the low density
($r_{s}\to\infty$) regime. 
%
\begin{figure}[t]
\begin{center}
\includegraphics[width=0.7\columnwidth]{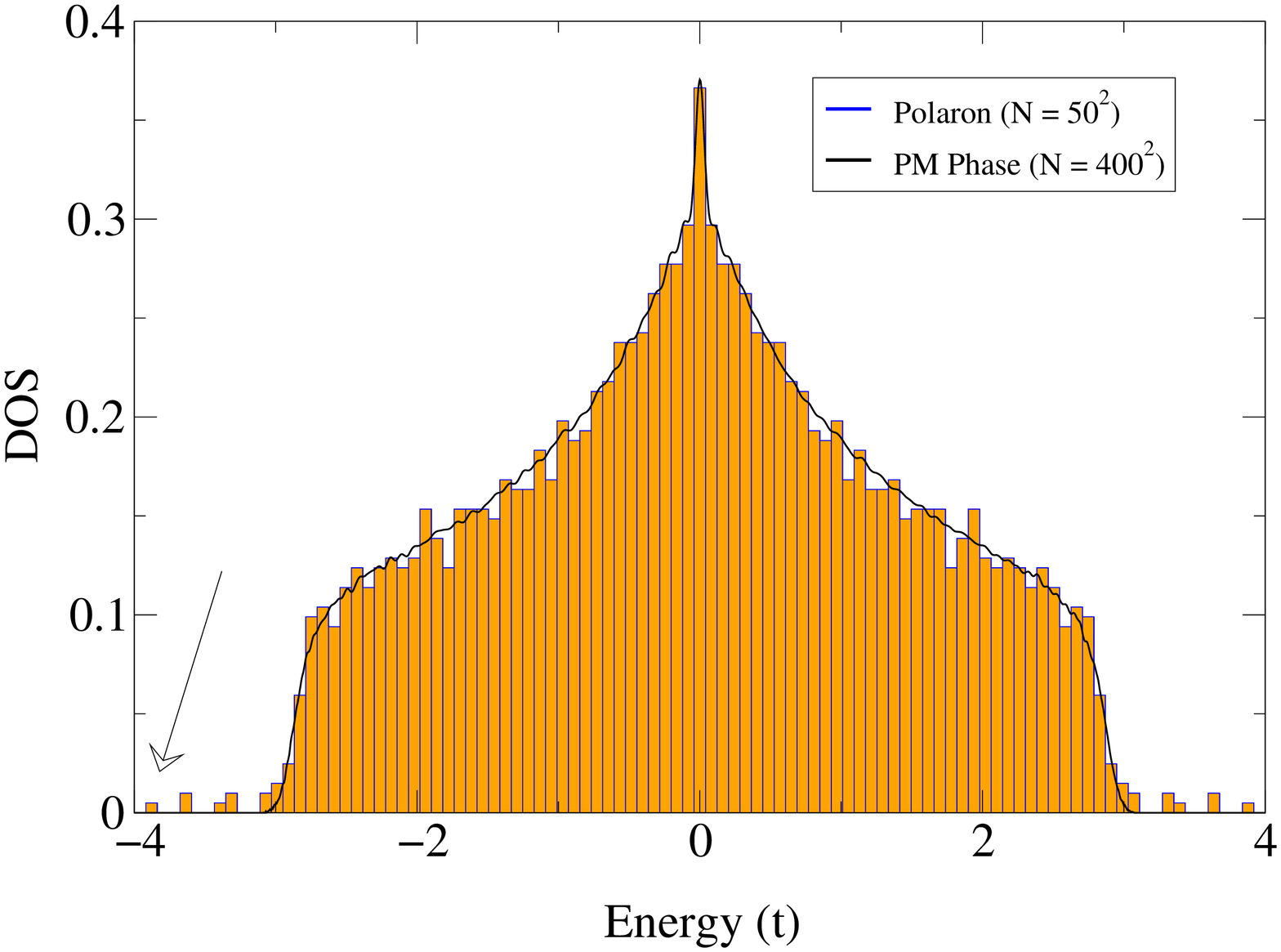}
\includegraphics[width=0.6\columnwidth]{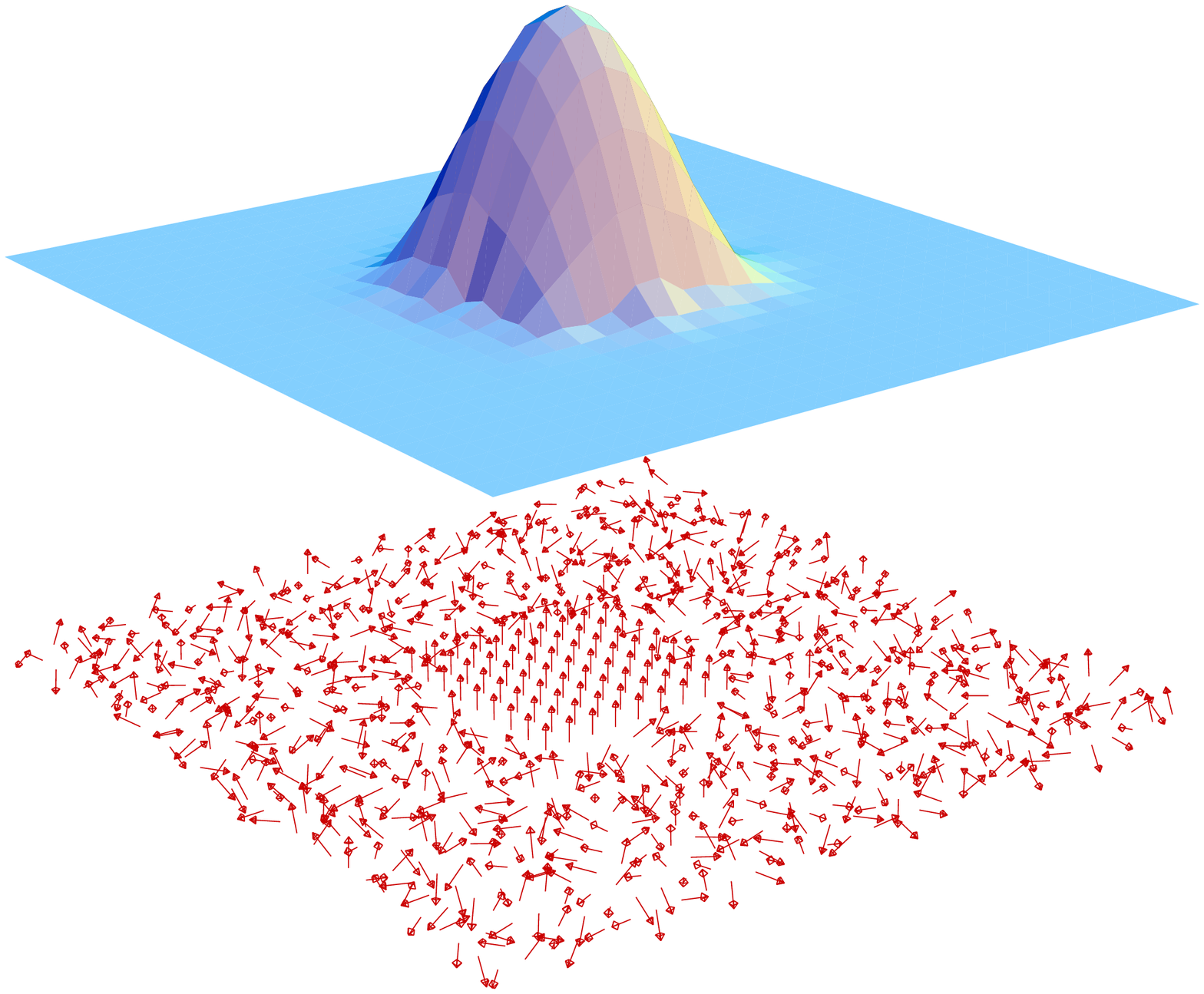}
\end{center}
\caption{Top: DOS for the 2-D
spin configuration shown on the bottom (exact diagonalization, bars)
and averaged DOS for the homogeneous case (recursive method, continuous
line). The magnetic polaron level (marked by an
arrow) agrees with $-4t\cos[\pi/(R+1)]$.
Bottom: Magnetic polaron wavefunction (top) and its
underlying lattice spin configuration (bottom).
\label{fig:Polaron_3D} }
\end{figure}
%
It is well known since Wigner \cite{Wigner:1934}
that, under such circumstances, the electrons arrange themselves
in a regular lattice.
Another aspect related to the reduced carrier density is precisely
the stability of free magnetic polarons. Within the simple DEM,
the only possibility for the stabilization of magnetic polarons is
at low electronic densities because, the lattice spins of several neighboring
unit cells are expected to participate in the magnetization cloud
of each electron. Were it otherwise, the electronic wave functions
would overlap considerably, destroying the polaron picture. On the
other hand, if a Wigner crystal is realizable in a system described
by eq.~(\ref{eq:H_DE_Coulomb}) then, as in any solid, there is
zero point motion of the electrons about their equilibrium
positions. If the electron itinerates among several unit cells during
this motion, the magnetic coupling, $J_{H}$, leads to the local
polarization of the lattice spins, producing a \textit{bound}
magnetic polaron. This process is obviously limited by the melting
of the electronic solid and, thus, the question arises of how to describe
these two tendencies for the polaron formation and the interplay of
Coulomb and magnetic interactions. 
A final aspect that cannot be neglected in this regime
is the fact that the classical approximation
for the spins implies an intrinsically
disordered electronic problem, at any non-zero temperature (a simple consequence of
the dependence of $t_{ij}$ upon the local spin texture). 
This requires, for instance, the calculation of the spectral properties for a disordered electronic system, executed here by means of the recursion method \cite{Haydock:1972} and exact diagonalization.


{\it Single Particle Regime}. At zero thermodynamic density 
the Coulomb interaction can be disregarded, and the problem
reduces to a single magnetic polaron. 
We consider the  electronic density of states
(DOS) for a given spin configuration, $\{{\bf S}_{i}\}$, generated
with a Boltzmann weight that includes the mean-field required to obtain
a given average magnetization, $\mathcal{M}$, of the lattice spins
\cite{Pereira:2004a}. Averaging over several realizations of disorder, 
we can write the electronic energy as: 
\begin{equation}
E_{el}(\mathcal{M},n_{e})=
\int\Theta\left(E_{F}(n_{e},\mathcal{M})-\epsilon\right)\epsilon\left\langle
  N(\epsilon,\mathcal{M})\right\rangle d\epsilon,
\label{eq:Eel_Full}
\end{equation}
where the explicit dependence of both the DOS ($N(\epsilon)$) and
Fermi energy ($E_{F}$) on the magnetization is emphasized. 
For $n_{e}\ll1$ we approximate eq.~(\ref{eq:Eel_Full}) simply 
by $E_{el}(\mathcal{M},n_{e})\approx E_{b}(\mathcal{M})n_{e}$,
with $E_{b}$ representing the bottom of the band that, in 3-D, is
found to lie at $-4t$ for the paramagnet ($\mathcal{M}=0$), and at $-6t$ in
the fully polarized ferromagnetic phase ($\mathcal{M}=1$).
Let us assume that the polaron consists of a cubic volume
of side $R$ (in units of the lattice parameter, $a$), inside which
$\mathcal{M}=1$. 
Outside the polaron the system remains paramagnetic ($\mathcal{M}=0$).
In Fig.~\ref{fig:Polaron_3D} we show exact diagonalization
results of a two-dimensional ferromagnetic region embedded in a
paramagnetic phase, leading to the formation of a bound state,
that is, a magnetic polaron \cite{next}. 
The free energy per electron, at a temperature $T$,
required to form a magnetic polaron from the paramagnetic
phase is given by:
\begin{eqnarray}
\Delta F_{\textrm{Pol}}\left(R,T\right)/n_e = 
 4t - 6t\cos\left(\pi/(R+1)\right) 
\nonumber 
\\
 + T R^{3}\log\left(2S+1\right) - T \mathcal{S}_{\textrm{Cfg}}(n_{e},R),
\label{eq:DF_With_Conf_Entropy}
\end{eqnarray}
where $\mathcal{S}_{\textrm{Cfg}}$ is the configurational entropy. 
The first three terms in (\ref{eq:DF_With_Conf_Entropy})
reflect the two competing effects at play: the first is the electron's
preference for a ferromagnetic background, accompanied by an energy
cost for localization; the second is the reduction of entropy
caused by the appearance of the (fully polarized) magnetic polarons.
For $n_e \ll 1$ we find that 
$\mathcal{S}_{\textrm{Cfg}}(n_{e},R)$ is always negligible \cite{next}. 

Minimizing eq.~(\ref{eq:DF_With_Conf_Entropy}) ($R\gg1$) with respect to $R$
gives:
\begin{equation}
R_{\textrm{eq}}\left(T\right)\simeq\left[2t\pi^{2}/(T\log\left(2S+1\right))\right]^{1/5}
\label{eq:R_eq}
\end{equation}
for the equilibrium radius of the magnetic polaron, increasing at low 
temperatures as $T^{-1/5}$. The stability condition for the
polaron is $\Delta F_{\textrm{Pol}}\left(R_{\textrm{eq}},T\right)<0$,
which provides the temperature, $T_{m}$,
below which the polarons first appear:
\begin{equation}
T_{m}/t = 8 \sqrt{2}/(25 \sqrt{5} \pi^{3}\log\left(2S+1\right)) \,.
\label{eq:Tm}
\end{equation}
Note that at $T=T_m$ the polarons are formed with a size given by
$R_{\textrm{eq}}(T_m)$, and hence, the crossover from the paramagnetic 
to the polaronic phase is discontinuous.

The previous results can also provide an estimate of the temperature
below which a homogeneous ferromagnetic state is established. Assume
that each electron forms a polaron that is 
randomly distributed over the system. 
Since the polaron radius increases according to (\ref{eq:R_eq}) as
the temperature is lowered, there is a critical temperature, 
$T_C^P$, below which the polarons {\it percolate} into a 
ferromagnetic phase. This percolation threshold is defined
from the criterion $n_{e}R^3_{\textrm{eq}} \simeq p_{c}$ ($p_c \approx 0.31$ for a cubic lattice), yielding from (\ref{eq:R_eq}) a Curie temperature,
\begin{equation}
T_{C}^{P}/t \simeq 2\pi^{2}/[\log\left(2S+1\right)]
\left(n_{e}/p_{c}\right)^{5/3} \, .
\label{eq:Tc_Percol}
\end{equation}
This temperature can be compared with the Curie temperature obtained
by a mean-field approach : $T^{MF}_C/t = 5 (1+S)/(3 S) n_e$. 
Notice that, for $n_e \to 0$, we have $T_C^P \ll T_C^{MF}$, indicating that the
mean-field calculation overestimates the Curie temperature in the
low density regime. 
For $T_{C}<T<T_{m}$, an anomaly in the paramagnetic susceptibility
is expected to signal the polaron presence, through an enhanced
effective moment. Naively, the transition from the polaronic phase to
the ferromagnetic phase would be first order, for the magnetization
jumps at $T_C$. Nevertheless, because this transition is percolative
in nature, one would expect a second order phase transition, in the
sense that the average magnetization should be weighted by the ``mass
of the infinite cluster'', which evolves continuously from the
percolation threshold.  


{\it Polaronic Wigner Crystal}. The previous estimate
of the Curie temperature for a small but finite density of electrons
has a fundamental problem: the non-interacting 
DEM is unstable towards phase separation at low densities
\cite{Nagaev:1998,Yunoki:1998,Arovas:1999,Kagan:1999,Alonso:2001a,Alonso:2001b}.
In the presence of Coulomb interactions, however, the phase separation
is frustrated by the large electrostatic energy price of confining
charge to a given region of the system. It is found that, for reasonable values of the
dielectric constant, the tendency for charge neutrality, which favors
small radius of electron rich regions, and the kinetic energy of
localization, lead to a strong suppression of the phase separation
region,  in temperature and electron concentration, and its replacement
by a polaronic phase \cite{next}.

At the lowest densities
($r_s \gg 1$) the Coulomb interaction is unscreened and larger than the 
kinetic energy in (\ref{eq:H_DE_Coulomb}). In order to minimize the Coulomb
energy, the electrons form a Wigner crystal and 
undergo zero point motion around their equilibrium position. Delocalization
of the electron wavefunction due to quantum fluctuations leads to the
polarization of the local spins around the crystalline positions.
The local polarization, exactly as in the case of the single polaron 
discussed previously, produces an extra confining potential to the electron.
Following the Wigner-Seitz approximation \cite{Pines:1963}, the 
Wigner crystal unit cell (much larger than the original
lattice spacing $a$) is approximated by an electrically neutral
spherical volume, inside which the ionic charge density is homogeneous. 
The electrostatic potential energy then
depends only upon $r$: the distance of the electron from the center
of the cell. The Hamiltonian for an electron in this
uniform charge and spin background is then:
\begin{equation}
\mathcal{H}_{W} = -6 t - \frac{3}{2}\frac{e^{2}}{r_{o}} 
+ \frac{p^{2}}{2m}+\frac{1}{2}m(\omega_{e}^{2} + \omega^2) r^{2}\,,
\label{eq:H_WignerCrystal}
\end{equation}
where $p$ is the electron momentum,  $m=1/(2 a^{2} t)$ the 
effective electron mass, and
$\omega_{e}^{2}=\omega_{p}^{2}/3=e^{2}/mr_{o}^{3}$, where $\omega_p$ is the
plasma frequency. In Eq.(\ref{eq:H_WignerCrystal}) $\omega$ is the frequency
of the confining potential due to the DEM mechanism, and is a variational
parameter. The radius of the magnetic polaron in the Wigner crystal
relates to $\omega$ by: 
\begin{equation}
R=\sqrt{3t/ \Omega} \, ,
\label{eq:R_vs_Omega}
\end{equation}
where $\Omega = \sqrt{\omega_e^2 + \omega^2}$ is the total
frequency of oscillation of the electron. Notice that the ground
state energy of (\ref{eq:H_WignerCrystal}) is $E_0 = -6 t - 3 e^2/(2 r_o)
+ 3   \Omega/2$ and hence the relative gain in free energy is:
\begin{equation}
\Delta\mathcal{F}_{WP}=-2t+3 \left(\Omega-\omega_{e}\right)/2+4 \pi
R^{3}T\log\left(2S+1\right)/3.
\label{eq:DF_WP}
\end{equation}
Minimization of eq.~(\ref{eq:DF_WP}) with respect to $\omega$ gives:
\begin{equation}
R(T)=\left\{ \begin{array}{c}
\begin{array}{ll}
R_S \, \, \left(T^*/T\right)^{1/5} & ,T>T^{*}\\
R_S & ,T \le T^{*}\end{array}\end{array}\right. \, , 
\label{eq:Req_WC_2}
\end{equation}
where $R_S = [3t/(  \omega_e)]^{1/2}$ is the saturation radius, and
$T^{*}/t= 9/[4\pi R_S^5 \log\left(2S+1\right)]$ is the temperature below
which the polaron radius saturates due to the interplay between the DEM
and the Coulomb interaction.


{\it Wigner Crystal Melting}. 
It is clear that the previous results are valid for temperatures so low as not
to melt the Wigner solid. The calculations for the independent polaron
model reveal that the temperatures for polaron stability are already
typically small for reasonable values of $t$, but the electronic
solid is much more sensitive to the temperature. The Wigner crystal 
melting temperature, $T_M$, can be estimated from the Lindemann's criteria
\cite{Jones:1996}: $T_M \approx 0.01 (e^2/a) n_e^{1/3}$. 
It is known from several numerical calculations
\cite{Candido:2004,Ortiz:1999,Jones:1996} 
on the stability of the one component plasma, that the maximum densities
and temperatures at which the Wigner crystal can exist correspond
to $r_{s}\sim50\textrm{-}100$, and $T\sim10\,\textrm{K}$. Values
of $r_{s}\sim50\textrm{-}100$ correspond to $n_{e}\sim10^{-6}\textrm{-}10^{-5}$
for  $t=1\,\textrm{eV}$ and $a=4\,\textrm{\AA}$. 
Because of the absence of magnetic interactions
between different polarons, the Wigner crystal is a superparamagnet. 
In the presence of other long-range
interactions (such as dipole-dipole) the polaronic Wigner crystal can exibith long range magnetic order. Increasing the electron density at $T=0$ causes the
Wigner crystal to quantum melt at some critical density with two
possible outcomes: a paramagnetic polaronic Fermi liquid or a
fully polarized ferromagnet. In both cases the carriers are mobile
and can screen the long-range part of the Coulomb interaction leading
to a Fermi liquid state. 
At finite temperatures, where the electron state cannot be described by the
zero point motions implicit in eq.~(\ref{eq:H_WignerCrystal})
alone, the crystal should follow the features of the phase diagram
for the electron gas \cite{Jones:1996}. The characterization of the
system in the neighborhood of the melting point, where the presence
of a polaron liquid is plausible (Fig.~\ref{fig:PhDiag_WS_Polarons}), 
is restrictively hard, even for the simple electron gas. Far from
this region, where the electron density is high enough to make the
screening process effective, one expects to retrieve the behavior
obtained before within the independent polaron model discussed previously.
Our results are summarized in the phase diagram presented in 
Fig.~\ref{fig:PhDiag_WS_Polarons}.


{\it Application to EuB$_6$}. While we have produced the general
phase diagram for the DEM model at low densities, it is interesting
to apply our results to a real system. EuB$_{6}$ is a good magnetic 
metal with extremely reduced electron density \cite{Sullow:1998Paschen:2000},
exhibiting all the characteristic signatures of a polaronic phase in Raman 
scattering \cite{Snow:2001}. Such experiments reveal that
the FM transition at $T^{{\rm exp}}_{C}\simeq15\,\textrm{K}$ is preceded by an
interval of temperatures ($T < T_m^{{\rm exp}} \approx 30 K$) 
where the system is dominated by the presence
of magnetic polarons. We have recently proposed that
EuB$_{6}$ is a DEM material in the low density regime
\cite{Pereira:2004a},
characterized by a hopping integral $t=0.55\,\textrm{eV}$, and a
carrier density per unit cell $n_{e}\sim10^{-3}$ (which, according 
to Fig.~\ref{fig:PhDiag_WS_Polarons}, puts EuB$_6$ far away from the WS) . 
If we incorporate such values in our previous results (see
Fig.~\ref{fig:PhDiag_WS_Polarons})  
we find $T_m \approx 43$ K and
$T^P_C \approx 15$ K, in accordance with the experimental data (notice
that the mean-field theory predicts much higher values for $T_C^{MF}$, in complete
disagreement with the experiments). These results show that
our description captures the polaron physics in EuB$_{6}$. 


{\it Conclusions}. We have studied the DEM at low densities in the
presence of Coulomb interactions and shown that its phase diagram is
very rich. We find
that at extreme low densities the system does not phase separate but,
instead, forms a Wigner crystal of magnetic polarons.  Due
to thermal or quantum effects, this crystal melts into a polaronic
Fermi liquid or a metallic ferromagnet. We find that the polaronic
liquid is stable in an intermediate temperature range $T^P_C < T <
T_m$. Above $T_m$ the polaronic liquid evolves into  
a paramagnetic metal while at low temperatures, $T<T^P_C$, 
it percolates into a 
metallic ferromagnet. Our theory has only two free parameters, the
number of electrons per site, $n_e$, and the hopping energy $t$,
and should apply to a broad class of materials that are described
by the diluted DEM. In particular, our results provide a good description of
the physics observed experimentally in EuB$_6$.


\acknowledgments
We thank L.~Brey, L.~De Giorgi, F.~Guinea, and N.~Peres, for 
illuminating discussions.  V.M.P., and J.M.B.L.S. are financed
by FCT and the E.U., through POCTI(QCA III). V.M.P. further
acknowledges Boston University for the hospitality and the financial
support of FCT, through grant ref. SFRH/BD/4655/2001. A.H.C.N. was
partially supported through NSF grant DMR-0343790.



\end{document}